\documentclass[final]{svjour3}
\usepackage{graphicx}
\usepackage{rotating}
\usepackage{amssymb}
\usepackage{mathptmx}
\usepackage[numbers,sort&compress]{natbib}
\usepackage{lipsum}
\usepackage{subfig}
\usepackage{sidecap}
\usepackage{wasysym}
\usepackage[font={small}]{caption}
\usepackage{bm}
\usepackage{tabularx}
\usepackage[hang,flushmargin,bottom]{footmisc} 
\makeatletter
\newcommand\subparagraph{%
  \@startsection{subparagraph}{5}
  {\parindent}
  {3.25ex \@plus 1ex \@minus .2ex}
  {-1em}
  {\normalfont\normalsize\bfseries}}
\makeatother
\makeatletter
\def\@maketitle{\newpage
\normalfont
\vbox to0pt{\if@twocolumn\vskip-39pt\else\vskip-49pt\fi
\nointerlineskip
\makeheadbox\vss}\nointerlineskip
\vbox to 0pt{\offinterlineskip\rubricwidth=\columnwidth
\vskip-12.5pt
\if@twocolumn\else 
   \divide\rubricwidth by144\multiply\rubricwidth by89 
   \vskip-\topskip
\fi
\hrule\@height0.35mm\noindent
\advance\fboxsep by.25mm
\global\advance\rubricwidth by0pt
\rubric
\vss}\vskip-42pt 
\if@twocolumn\else
 \gdef\footnoterule{%
  \kern-3\p@
  \hrule\@width\columnwidth  
  \kern2.6\p@}
\fi
 \setbox\authrun=\vbox\bgroup
     \hrule\@height 9mm\@width0\p@
     \pretolerance=10000
     \rightskip=0pt plus 4cm
    \nothanksmarks
    {\authorfont
    \setbox0=\vbox{\setcounter{auth}{1}\def\and{\stepcounter{auth} }%
                   \hfuzz=2\textwidth\def\thanks##1{}\@author}%
    \setcounter{footnote}{0}%
    \global\value{inst}=\value{auth}%
    \setcounter{auth}{1}%
    \if@twocolumn
       \rightskip43mm plus 4cm minus 3mm
    \else 
    \fi
\def\and{\unskip\nobreak\enskip{\boldmath$\cdot$}\enskip\ignorespaces}%
    \noindent\ignorespaces\@author\vskip3.62pt} 
    {\LARGE\bfseries
     \noindent\ignorespaces
     \@title \par}\vskip 1.62pt\relax 
    \if!\@subtitle!\else
      {\large\bfseries
      \pretolerance=10000
      \rightskip=0pt plus 3cm
      \vskip-5pt
      \noindent\ignorespaces\@subtitle \par}\vskip 0.24pt
    \fi
    \small
    \if!\@dedic!\else
       \par
       \normalsize\it
       \addvspace\baselineskip
       \noindent\@dedic
    \fi
 \egroup 
 \@tempdima=\headerboxheight
 \advance\@tempdima by-\ht\authrun
 \unvbox\authrun
 \ifdim\@tempdima>0pt
    \vrule width0pt height\@tempdima\par
 \fi
 \noindent{\small\@date\vskip -2.24pt} 
 \global\@minipagetrue
 \global\everypar{\global\@minipagefalse\global\everypar{}}%
}
\makeatother
\usepackage{titlesec}
\let\subparagraph\relax 
\usepackage{titlesec}
\makeatletter
\journalname{Journal of Low Temperature Physics}
\bibpunct{}{}{,}{s}{}{,}

\newcommand{\EE}{\emph{E\kern0.75ptE} }
\newcommand{\TT}{\emph{T\kern0.75ptT} }
\newcommand{\BB}{\emph{B\kern0.75ptB} }

\newcommand{\spider}{\textsc{Spider} }

\begin{document}

\newcommand{\hdblarrow}{H\makebox[0.9ex][l]{$\downdownarrows$}-}
\title{280 GHz Focal Plane Unit Design and Characterization for the \textsc{Spider}-2 Suborbital Polarimeter}

\author{A.S.~Bergman\textsuperscript{1} \and P.A.R.~Ade\textsuperscript{2} \and S.~Akers\textsuperscript{3} \and M.~Amiri\textsuperscript{4} \and J.A.~Austermann\textsuperscript{5} \and J.A.~Beall\textsuperscript{5} \and D.T.~Becker\textsuperscript{5} \and S.J.~Benton\textsuperscript{1} \and J.J.~Bock\textsuperscript{6,7} \and J.R.~Bond\textsuperscript{8} \and S.A.~Bryan\textsuperscript{9} \and H.C.~Chiang\textsuperscript{10,11} \and C.R.~Contaldi\textsuperscript{12} \and R.S~Domagalski\textsuperscript{13} \and O.~Dor\'e\textsuperscript{6,7} \and S.M.~Duff\textsuperscript{5} \and A.J.~Duivenvoorden\textsuperscript{14} \and H.K.~Eriksen\textsuperscript{15} \and M.~Farhang\textsuperscript{8,13} \and J.P.~Filippini\textsuperscript{16,17} \and L.M.~Fissel\textsuperscript{18,13} \and A.A.~Fraisse\textsuperscript{1} \and K.~Freese\textsuperscript{19,14} \and M.~Galloway\textsuperscript{20} \and A.E.~Gambrel\textsuperscript{1} \and N.N.~Gandilo\textsuperscript{21,22} \and K.~Ganga\textsuperscript{23} \and A.~Grigorian\textsuperscript{5} \and R.~Gualtieri\textsuperscript{16} \and J.E.~Gudmundsson\textsuperscript{14} \and M.~Halpern\textsuperscript{4} \and J.~Hartley\textsuperscript{20} \and M.~Hasselfield\textsuperscript{24} \and G.~Hilton\textsuperscript{5} \and W.~Holmes\textsuperscript{7} \and V.V.~Hristov\textsuperscript{6} \and Z.~Huang\textsuperscript{8} \and J.~Hubmayr\textsuperscript{5} \and K.D.~Irwin\textsuperscript{25,26} \and W.C.~Jones\textsuperscript{1} \and A.~Khan\textsuperscript{16} \and C.L.~Kuo\textsuperscript{25} \and Z.D.~Kermish\textsuperscript{1} \and S.~Li\textsuperscript{1,13,27} \and P.V.~Mason\textsuperscript{6} \and K.~Megerian\textsuperscript{7} \and L.~Moncelsi\textsuperscript{6} \and T.A.~Morford\textsuperscript{6} \and J.M.~Nagy\textsuperscript{3,28} \and C.B.~Netterfield\textsuperscript{13,20} \and M.~Nolta\textsuperscript{8} \and B.~Osherson\textsuperscript{16} \and I.L.~Padilla\textsuperscript{13,21} \and B.~Racine\textsuperscript{15,29} \and A.S.~Rahlin\textsuperscript{30,31} \and S.~Redmond\textsuperscript{32} \and C.~Reintsema\textsuperscript{5} \and L.J.~Romualdez\textsuperscript{32} \and J.E.~Ruhl\textsuperscript{3} \and M.C.~Runyan\textsuperscript{7} \and T.M.~Ruud\textsuperscript{15} \and J.A.~Shariff\textsuperscript{8} \and E.C.~Shaw\textsuperscript{16} \and C.~Shiu\textsuperscript{1} \and J.D.~Soler\textsuperscript{33} \and X.~Song\textsuperscript{1} \and A.~Trangsrud\textsuperscript{6,7} \and C.~Tucker\textsuperscript{2} \and R.S.~Tucker\textsuperscript{6} \and A.D.~Turner\textsuperscript{7} \and J.~Ullom\textsuperscript{5} \and J.F.~van der List\textsuperscript{1} \and J.~Van Lanen\textsuperscript{5} \and M.R.~Vissers\textsuperscript{5} \and A.C.~Weber\textsuperscript{7} \and I.K.~Wehus\textsuperscript{15} \and S.~Wen\textsuperscript{3} \and D.V.~Wiebe\textsuperscript{4} \and E.Y.~Young\textsuperscript{1}
}

\institute{\textsuperscript{1}Department of Physics, Princeton University, Princeton, NJ, USA\\
\textsuperscript{2}School of Physics and Astronomy, Cardiff University, UK\\
\textsuperscript{3}Physics Department, Center for Education and Research in Cosmology and Astrophysics, Case Western Reserve University, Cleveland, OH, USA \\
\textsuperscript{4}Department of Physics and Astronomy, University of British Columbia, Vancouver, BC, Canada\\
\textsuperscript{5}National Institute of Standards and Technology, Boulder, CO, USA\\
\textsuperscript{6}Division of Physics, Mathematics and Astronomy, California Institute of Technology, Pasadena, CA, USA\\
\textsuperscript{7}Jet Propulsion Laboratory, Pasadena, CA, USA \\
\textsuperscript{8}Canadian Institute for Theoretical Astrophysics, University of Toronto, Toronto, ON, Canada\\
\textsuperscript{9}School of Earth and Space Exploration, Arizona State University, Tempe, AZ, USA \\
\textsuperscript{10}School of Mathematics, Statistics and Computer Science, University of KwaZulu-Natal, Durban, South Africa \\
\textsuperscript{11}National Institute for Theoretical Physics (NITheP), KwaZulu-Natal, South Africa \\
\textsuperscript{12}Blackett Laboratory, Imperial College London, SW7 2AZ, London, UK \\
\textsuperscript{13}Department of Astronomy and Astrophysics, University of Toronto, Toronto, ON, Canada\\
\textsuperscript{14}The Oskar Klein Centre for Cosmoparticle Physics, Department of Physics, Stockholm University, Stockholm, Sweden\\ 
\textsuperscript{15}Institute of Theoretical Astrophysics, University of Oslo, Oslo, Norway \\
\textsuperscript{16}Department of Physics, University of Illinois at Urbana-Champaign, Urbana, IL, USA\\
\textsuperscript{17}Department of Astronomy, University of Illinois at Urbana-Champaign, Urbana, IL, USA\\
\textsuperscript{18}National Radio Astronomy Observatory, Charlottesville, NC, USA\\
\textsuperscript{19}Department of Physics, University of Michigan, Ann Arbor, MI, USA\\
\textsuperscript{20}Department of Physics, University of Toronto, Toronto, ON, Canada\\
\textsuperscript{21}Department of Physics and Astronomy, Johns Hopkins University, Baltimore, MD, USA\\
\textsuperscript{22}NASA Goddard Space Flight Center, Greenbelt, MD, USA\\
\textsuperscript{23}APC, Univ. Paris Diderot, CNRS/IN2P3, CEA/Irfu, Obs de Paris, Sorbonne Paris Cité, France\\
\textsuperscript{24}Pennsylvania State University, University Park, PA, USA\\
\textsuperscript{25}Department of Physics, Stanford University, Stanford, CA, USA\\
\textsuperscript{26}SLAC National Accelerator Laboratory, Menlo Park, CA, USA \\
\textsuperscript{27}Department of Mechanical and Aerospace Engineering, Princeton University, Princeton, NJ, USA \\
\textsuperscript{28}Dunlap Institute for Astronomy \& Astrophysics, University of Toronto, Toronto, ON, Canada\\
\textsuperscript{29}Harvard-Smithsonian Center for Astrophysics, Cambridge, MA, USA\\
\textsuperscript{30}Fermi National Accelerator Laboratory, Batavia, IL, USA\\
\textsuperscript{31}Kavli Institute for Cosmological Physics, University of Chicago, Chicago, IL, USA\\
\textsuperscript{32}University of Toronto Institute for Aerospace Studies, Toronto, ON, Canada\\
\textsuperscript{33}Max Planck Institute for Astronomy, Heidelberg, Germany\\
\email{stevie@princeton.edu}}

\maketitle

\begin{abstract}
We describe the construction and characterization of the 280\,GHz bolometric focal plane units (FPUs) to be deployed on the second flight of the balloon-borne \spider instrument. 
These FPUs are vital to \textsc{Spider}'s primary science goal of detecting or placing an upper limit on the amplitude of the primordial gravitational wave signature in the cosmic microwave background (CMB) by constraining the $B$-mode contamination in the CMB from Galactic dust emission. 
Each 280\,GHz focal plane contains a $16 \times 16$  grid of corrugated silicon feedhorns coupled to an array of aluminum-manganese transition-edge sensor (TES) bolometers fabricated on 150\,mm diameter substrates. In total, the three 280\,GHz FPUs contain 1,530 polarization sensitive bolometers (765 spatial pixels) optimized for the low loading environment in flight and read out by time-division SQUID multiplexing. In this paper we describe the mechanical, thermal, and magnetic shielding architecture of the focal planes and present cryogenic measurements which characterize yield and the uniformity of several bolometer parameters. The assembled FPUs have high yields, with one array as high as 95\% including defects from wiring and readout. 
We demonstrate high uniformity in device parameters, finding the median saturation power for each TES array to be $\sim$3\,pW at 300\,mK with a less than 6\% variation across each array at $1\sigma$. These focal planes will be deployed alongside the 95 and 150\,GHz telescopes in the \textsc{Spider}-2 instrument, slated to fly from McMurdo Station in Antarctica in December 2018.
\end{abstract}


\section{Introduction}\label{sec:intro}\vspace{-.5em}
The field of observational cosmology is moving at a rapid pace, with recent measurements of the polarization in the cosmic microwave background radiation (CMB) pushing the frontier of our investigative capabilities. The current focus is on the primoridal $B$-mode polarization in the CMB, expected to have been sourced by gravitational waves passing through the surface of last scattering.\cite{ref:planck2015_fg} If this signal exists, it is obscured in observations by foreground signal, in particular that from Galactic dust.\cite{ref:bicep2016, ref:planck2015}

The \spider instrument includes six refracting telecopes designed to measure the polarization in the CMB from a ballooning platform. These telescopes are collectively encased within a 1300\,L liquid helium cryostat with an approximately 20\,day hold time. 
Over 2300 transition-edge sensors (TES) are cooled to their 300\,mK operating temperature by a closed-cycle, helium-3 adsorption refrigerator at the base of each telescope, and read out by time-division multiplexing (TDM) superconducting quantum interference devices (SQUID).\cite{ref:crill2008, ref:filippini2011, ref:fraisse2013, ref:rahlin2014, ref:gudmundsson2015}

The first \textsc{Spider} instrument was launched in January 2015 from McMurdo Station in Antarctica and contained six 95 and 150\,GHz slot-antenna-coupled TES arrays fabricated at JPL.\cite{ref:runyan2011, ref:spider12015} During the 16\,day flight, \textsc{Spider}-1 observed 10\% of the sky from an altitude of 36\,km and collected 2 TB of raw data, currently under analysis.\cite{ref:nagy2017}

The second flight is scheduled for December 2018 and will re-deploy three proven \textsc{Spider}-1 telescopes alongside three NIST-fabricated feedhorn-coupled 280\,GHz TES focal plane units (FPUs).\cite{ref:hubmayr2016} Otherwise, \textsc{Spider}-2 will have effectively the same construction as the first instrument. 

The 280\,GHz band has high signal-to-noise in Galactic dust foregrounds and complements the 95 and 150\,GHz data, enabling improved characterization of the foreground spectrum.
The continuum contribution of the atmosphere limits detector sensitivity at frequencies above about 200\,GHz for ground-based observations. At float altitudes the atmospheric loading is reduced by several orders of magnitude, enabling background-limited observations of the microwave sky at these frequencies.\cite{ref:fraisse2013}

In this proceeding we describe the mechanical, thermal, and magnetic shielding architecture of the 280\,GHz FPUs and present cryogenic measurements of the yield and bolometer parameters. 

\vspace{-5mm}
\section{280 GHz Focal Plane}\label{sec:280fpu}\vspace{-.5em}
The unique constraints of ballooning necessitate a compact, modular FPU structure (\S\ref{sec:sensor_pack}) and robust shielding from ambient magnetic fields at float (\S\ref{sec:mag_shielding}).\cite{ref:runyan2011, ref:odea2011}
The 280\,GHz FPUs were designed to utilize proven \textsc{Spider}-1 packaging and magnetic shielding with the NIST sensor array units (\S\ref{sec:sensor_stack}), and to fit within the \textsc{Spider}-1 optical\cite{ref:rahlin2014} and cryogenic design\cite{ref:gudmundsson2015} such that the telescope could slot in to any port in the cryostat. 

Figure~\ref{fig:y3} shows a fully assembled FPU, including the sensor array assembly centered on the copper faceplate on the sky side of the box. The first two stages of SQUID readout are under the sensors, within several layers of high-permeability (high-$\mu$) and superconducting shielding. The 280\,GHz \spider FPU packaging complies with the previously mentioned constraints while maintaining the advantages of the low loading environment at 36\,km altitude.

\vspace{-5mm}
\subsection{Sensor Array Assembly}\label{sec:sensor_stack}\vspace{-.5em}
The 280\,GHz sensor array assembly is composed of a $16 \times 16$ array of conical, corrugated feedhorns coupled to a monolithic detector array fabricated on a 150\,mm diameter silicon wafer. The detector array stack consists of the silicon backshort, detector array wafer, and feedhorn interface wafer. These components are pressed against the corrugated feedhorn array via a 300\,mK copper bracket and BeCu springs. The TES wafer is heat sunk to 300\,mK through the silicon backshort, which is gold-plated on the bracket side. The detector stack is thermally connected to the 300\,mK bracket via tack-bonded gold ribbons, seen in Figure~\ref{fig:sau_skyin}.

On the detector array, each feedhorn is coupled to two TESs, one for each orthogonal polarization mode. Each bolometer pair, considered one pixel, is rotated 45$^{\circ}$ from its neighbor for simultaneous coverage of Q and U Stokes parameters. Every TES bolometer island contains two sensors in series with differing superconducting critical temperatures ($T_c$): a 420\,mK $T_c$ aluminum-manganese (AlMn) sensor for flight operations and a 1.6\,K $T_c$ Al sensor for on the ground pixel characterization.\cite{ref:hubmayr2016, ref:duff2016, ref:mcmahon2009, ref:henning2010, ref:deiker2004}

In total, the three \textsc{Spider}-2 280\,GHz focal planes contain 765\,spatial pixels and 1,530 polarization sensitive bolometers. These exhibit an electrical noise equivalent power (NEP) $=2.6 \times 10^{-17} \mathrm{W}/\sqrt{\mathrm{Hz}}$, as reported in Hubmayr, et al, 2016. Details of the sensor array and detector stack, including corrugated feedhorns and pixel fabrication, are well-described in previous publications.\cite{ref:hubmayr2016, ref:duff2016} 

\vspace{-5mm}
\subsection{Sensor Packaging}\label{sec:sensor_pack}
\begin{figure}
  \centering
  \subfloat[]{\includegraphics[width=0.52\linewidth,keepaspectratio]{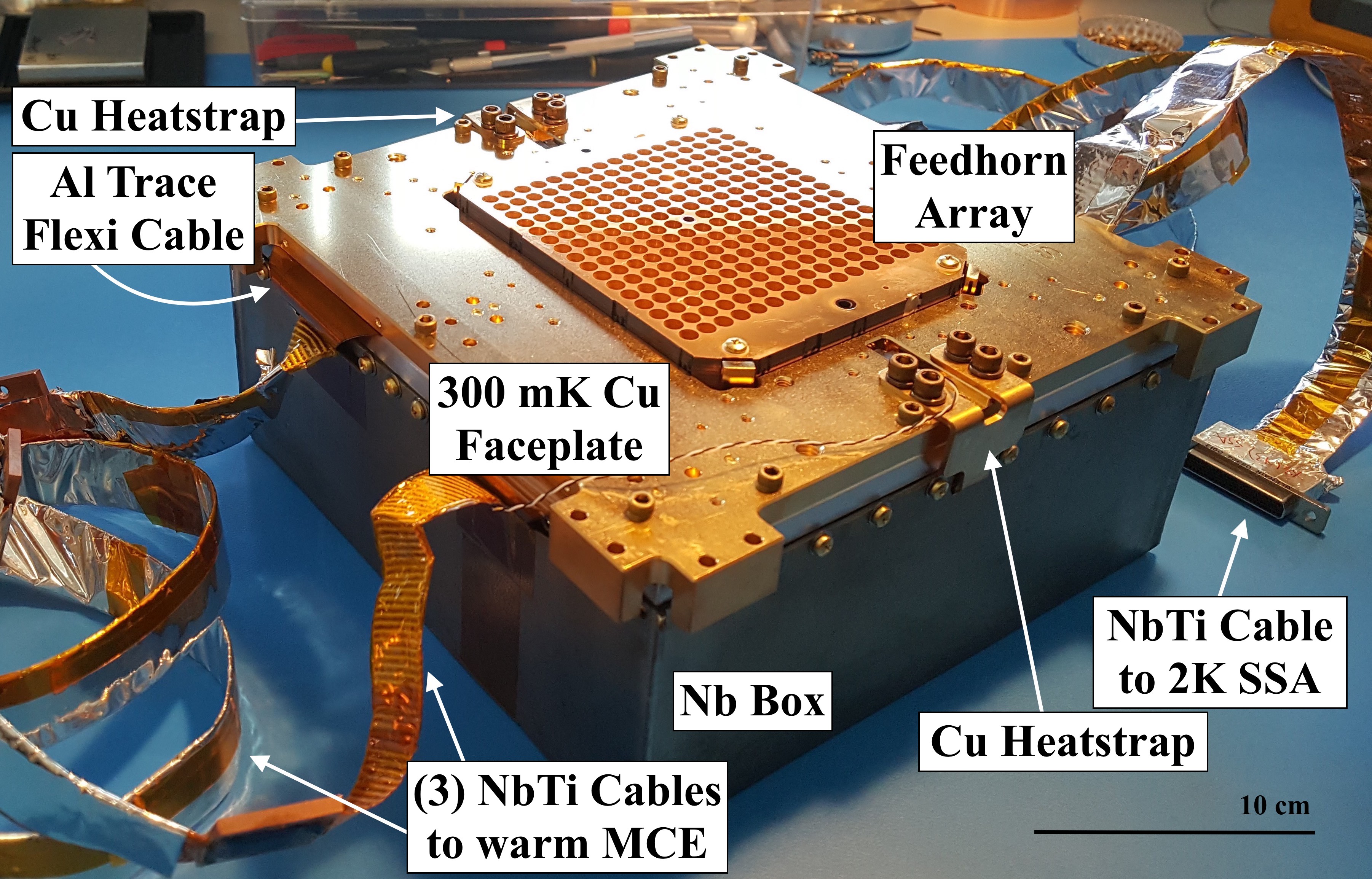}\label{fig:y3}}
  \hfill
  \subfloat[]{\includegraphics[width=0.47\textwidth]{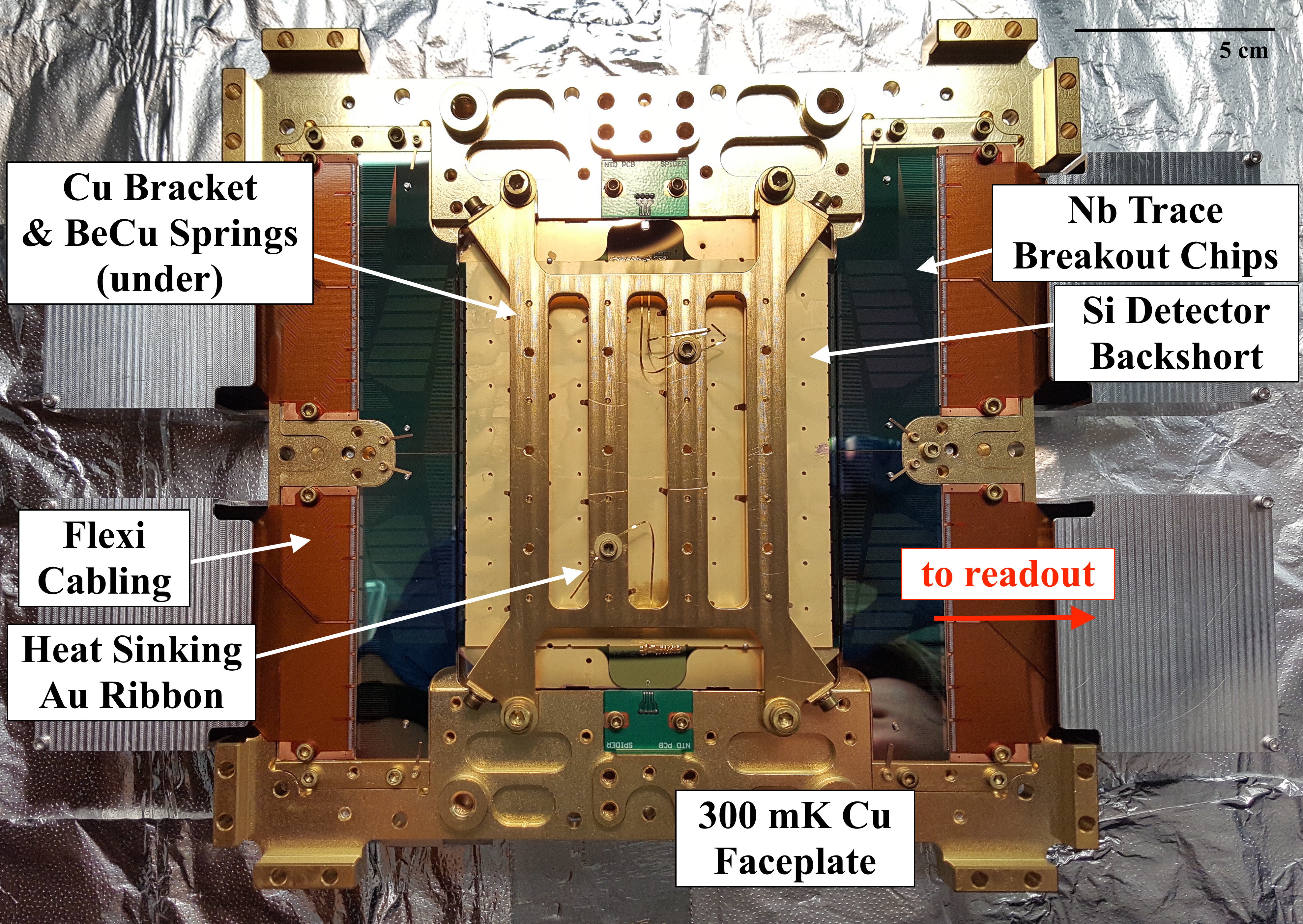}\label{fig:sau_skyin}}
  \caption{\footnotesize \textbf{(a)}: Fully assembled 280\,GHz focal plane unit, ready for installation. The \textsc{Spider}-2 FPUs are designed to reuse proven \textsc{Spider}-1 packaging and magnetic shielding, hybridized with the NIST sensor array units. Image (a) shows the 300\,mK gold-plated copper faceplate, with the corrugated feedhorns of the sensor array assembly set in the cutout at the center. The two gold-plated copper heat straps are thermally sunk to the faceplate and bent in to the box, cooling the internal components from the center to minimize trapping magnetic flux. The superconducting NbTi cabling electrically connects the first and second stage multiplexing SQUIDs to the 2K SSA and warm readout electronics. The niobium box serves as both the outer layer of the packaging and as superconducting magnetic shielding. \textbf{(b)}: Sensor array unit. Here, the  sky side of the detector stack is in to the page. This focal plane is in the middle of assembly, with detector and flexible cabling (``flexi") bondpads exposed in preparation for bonding. Nb-traced silicon breakout chips are used to mate the flexi cable and detector connections. For bonding, the flexi cables are protected by temporary covers, and the MUX readout PCBs are folded under the faceplate (between the table and faceplate in the photo).}
  \vspace{-1.5em}
\end{figure}

\vspace{-.5em}
Thermal connections on and within the focal plane begin with the gold-plated copper faceplate, which is itself thermally coupled to the cold point of the $3$He adsorption refrigerator at the base of the telescope via a copper bus and 300\,mK ring cradling the FPU. The high heat capacity faceplate and stainless steel mounting blocks between the ring and FPU serve as low-pass filters for any thermal fluctuations from the adsorption refrigerator. The components internal to the FPU are maintained at 300\,mK by two gold-plated copper heat straps. These are contorted such that they can be firmly heat sunk to the copper faceplate on the outside of the box, as shown in Figure~\ref{fig:y3}, then bent inside to form solid thermal contact with (1) the top of the niobium box and high-$\mu$ Cryoperm sheet and (2) the high-$\mu$ sleeves and copper backplates of the four multiplexing (MUX) SQUID readout boards. The thermal straps cool these components from the center to avoid trapping magnetic flux within the shielded focal plane box.\\
\indent The multiplexing SQUID readout is set behind the detector stack, within the shielding niobium and aluminum boxes, and connected to the sensor array unit via superconducting aluminum traced flexible circuits (``flexi cables"). The flexi cables electrically connect the TES bondpads to the silicon shunt resistor and MUX chips. At the detector end, the flexi cables are  secured on the underside of the faceplate as in Figure~\ref{fig:sau_skyin}. Niobium-traced silicon breakout chips (NIST-designed and fabricated) are used between the detector wafer and flexi cabling, due to the fact that the 170\,$\mu$m detector bondpad pitch could not be matched by commercial flexi cables.\\
\indent The shunt resistor and MUX chips are mounted on Tech-Etch printed circuit boards (PCBs) attached to the flexi cabling. These PCBs connect to superconducting NbTi cables via a breakout board, then exit the FPU through slits in the niobium box (see cabling exiting the FPU in Figure~\ref{fig:y3}). The NbTi cables connect the FPU to the 2K SQUID series array (SSA) amplifier and warm Multi-Channel Electronics mounted to hermetic flanges on the belly of the cryostat.\\
\begin{figure}[htbp]
\centering
  \includegraphics[width=.9\linewidth, keepaspectratio]{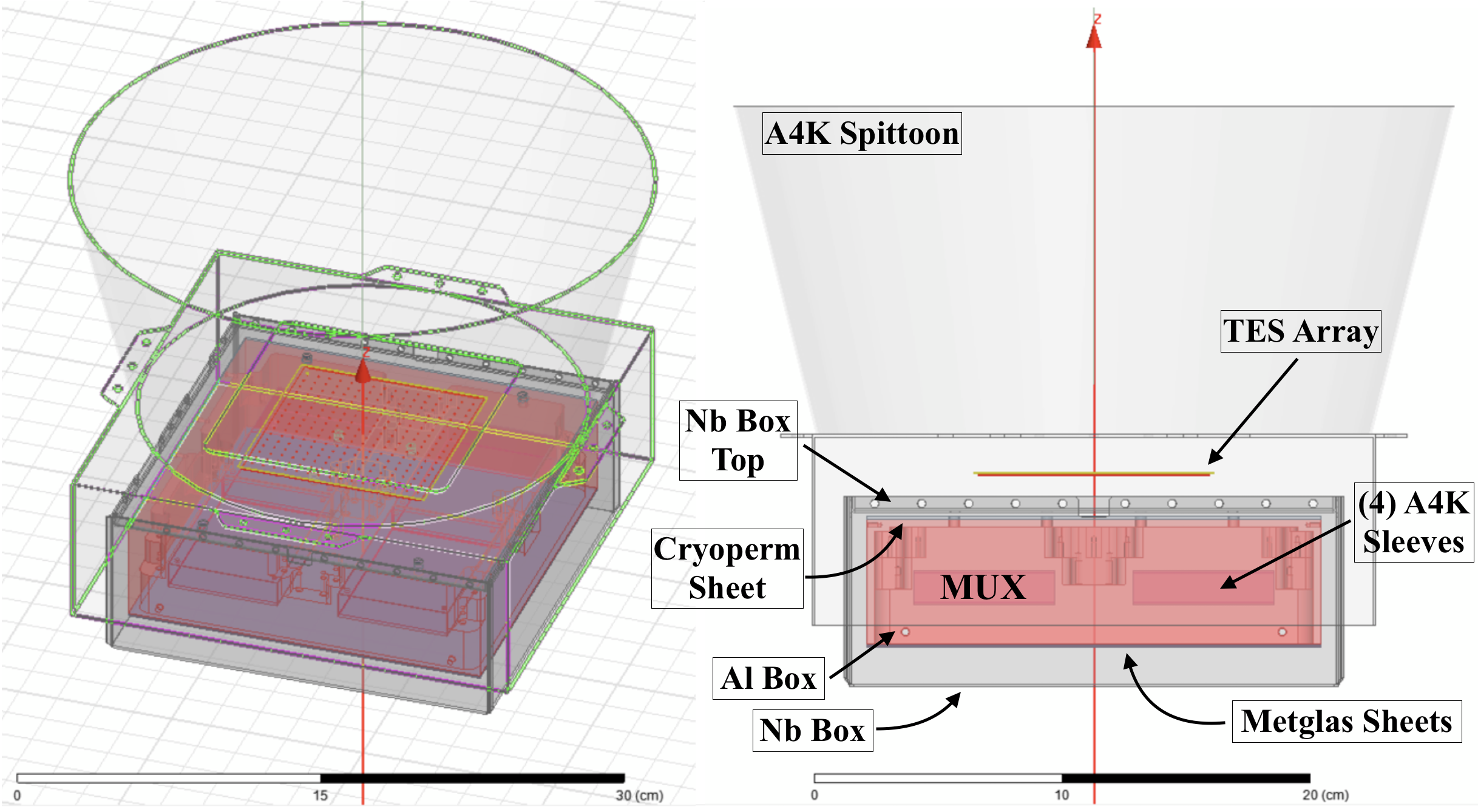}
\caption{\footnotesize Mulit-layer magnetic shielding. The four SQUID readout boards (one of which is labeled ``MUX" in the image) are shielded by high-permeability (high-$\mu$) A4K sleeves, a superconducting Al box that is closed at the bottom by high-$\mu$ Metglas, and a layer of high-$\mu$ Cryoperm above the Al box. That is all enclosed within a superconducting Nb box and topped by an A4K ``spittoon." In addition to the shielding shown here, the \textsc{Spider}-1 telescopes had two cylindrical high-$\mu$ shields along their full length (with the exception of one telescope with only one of these shields). This shielding was designed to attenuate ambient magnetic fields at the SQUIDs by $10^7$ and at the detector wafer by $10^4$.}
\label{fig:magshield}
\vspace{-1.5em}
\end{figure}

\vspace{-8mm}
\subsection{Magnetic Shielding}\label{sec:mag_shielding}\vspace{-.5em}
In ballooning, the magnetic environment of the payload is changing over time in a way that is challenging to measure or model. It is not easily removed in post-processing as part of scan synchronous noise, and will not necessarily integrate down over time.\cite{ref:stiehl2011} These time-varying magnetic fields can impact instrument performance by altering the critical temperature of the TES and by adding flux through the SQUIDs, both of which can mimic signal. 
At the SSA, a nonuniform field can cause destructive interference of the V-$\phi$ curves, resulting in weak signal amplification.\cite{ref:runyan2011, ref:odea2011}\\
\indent The SQUIDs within the \spider 280\,GHz FPU box are shielded by several layers of high-$\mu$ and superconducting metals. These include Amuneal-4K (A4K) high-$\mu$ sleeves, superconducting Al and Nb boxes, and high-$\mu$ Metglas sheets (Figure~\ref{fig:magshield}). This shielding was designed to attenuate ambient magnetic fields at the SQUIDs by $10^7$ and at the detector wafer by $10^4$. These values were deemed sufficient prior to the 2015 flight from laboratory testing and magnetic modeling.\citep{ref:runyan2011} With the exception of a Nb backshort found in the \textsc{Spider}-1 FPUs that served as both a detector backshort and Nb box top, the 280\,GHz magnetic shielding configuration is identical to the \textsc{Spider}-1 strategy described in Runyan, et al, 2011. From simulations using the Ansys Maxwell 3D magnetostatic software, we found that we can meet our shielding goals without the Nb backshort.\\ 
\indent In this configuration, the TES are outside the superconducting boxes, and thus are not as well shielded as the SQUIDs. This is addressed by placing a high-$\mu$ ``spittoon''  above the focal plane. The SSA is shielded separately from the rest of the components by a high-$\mu$ sleeve inside a superconducting Nb case, which is then wrapped in ten layers of Metglas sheeting. Finally, the entire insert is surrounded by two layers of A4K.\\
\indent Due to the non-uniformity in magnetic coupling across the wafer found in lab testing, dark SQUIDs (SQUIDs with no TES) can provide the magnitude of scan-synchronous pickup within the FPU box, but not an adequate template for noise removal. 
From analysis of \textsc{Spider}-1 flight data, the scan-synchronous noise in the dark SQUID channels is less than $1\%$ and can therefore be entirely accounted for by the estimation of cross-talk.\cite{ref:bicep2014_2, ref:dekorte2003} 
This indicates that the \textsc{Spider}-1 design provided sufficient magnetic shielding at the SQUIDs; thus we were motivated to utilize the same architecture for the \textsc{Spider}-2 focal planes. \\
\indent To compare the pickup in the \textsc{Spider}-1 and -2 focal planes, we took measurements of detector bias on transition with a .8\,G applied field, comparable to Earth's, for both a 95 and 280\,GHz FPU. We found there is not an appreciable difference in the magnitude of pickup in the devices when TES sensitivity is considered.\\
\indent Notably, the pickup is highest when the TESs are in their superconducting state; indicative of parasitic pickup in the wiring that connects the TESs to the SQUIDs, and not due to $T_c$ coupling to the magnetic field. These measurements showed no dark SQUID pickup in either FPU, further indication that the shielding inside the FPU box is sufficient.\\
\indent In the 280\,GHz FPU, this pickup in the wiring is approximately 10 times that of the dark SQUID noise floor; however these tests were done without the high-$\mu$ spittoon and multiple cylindrical shields. 
From simulation, these shields attenuate ambient fields by $10^3$, therefore we expect the full shielding configuration to mitigate this coupling below the noise floor.
\begin{figure}
  \centering
  \subfloat[]{\includegraphics[width=0.485\textwidth]{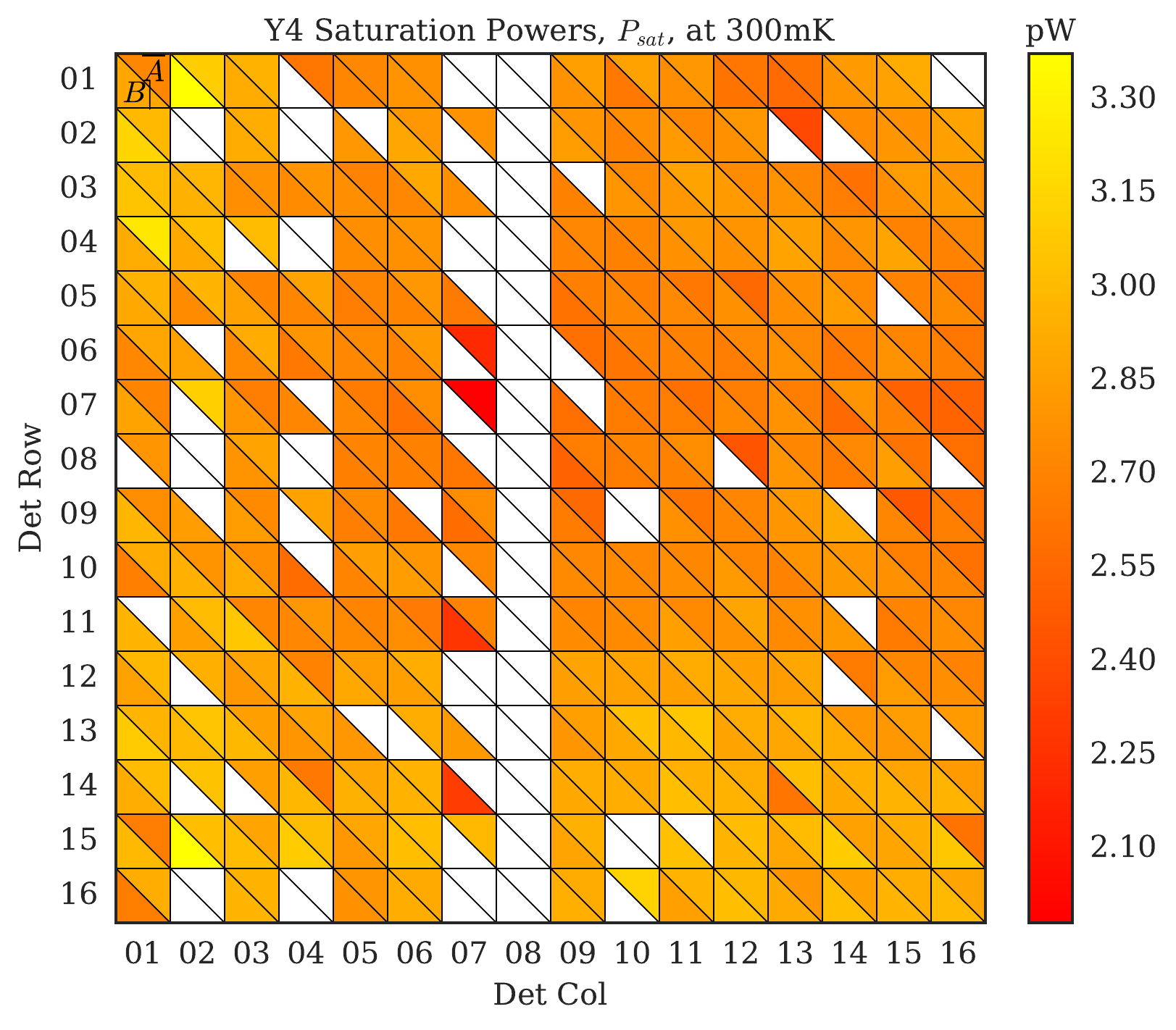}\label{fig:Psat_phys}}
    \hfill
  \subfloat[]{\includegraphics[width=0.5\textwidth]{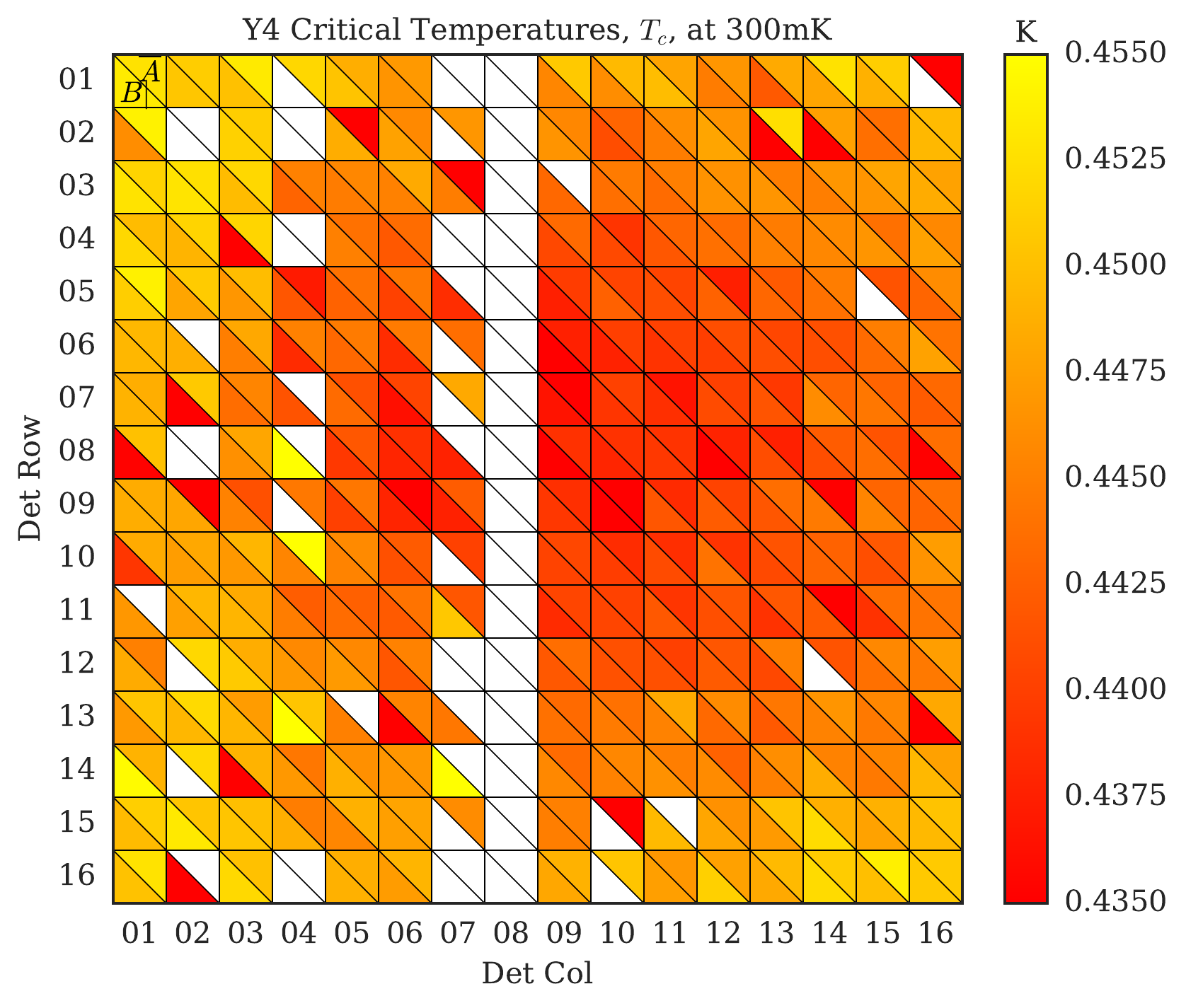}\label{fig:Tcrit_phys}}
  \caption{\footnotesize \textbf{(a)}: Saturation power across the detector wafer of one FPU (titled ``Y4"). The median $P_{sat}$ is 2.79\,pW, with a 5.6\% (.16\,pW) spread at 1$\sigma$ across the wafer. \textbf{(b)}: Per-TES critical temperature for the Y4 FPU. The median critical temperature is 445\,mK with a 1.6\%, or 7\,mK, spread at 1$\sigma$. A and B detectors, shown here as top (A) and bottom (B) halves of a pixel square, are orthogonal (\S\ref{sec:sensor_stack}). Note that white detetectors are dead (\S\ref{sec:characterization}), however detector column 8 has since been fixed.}
  \label{fig:phys_params}
  \vspace{-1.5em}
\end{figure}
\vspace{-2mm}
\section{Focal Plane Characterization and End-to-End Fabrication Yield}\label{sec:characterization}\vspace{-.5em}
Despite the elaborate packaging scheme described in \S\,\ref{sec:280fpu}, we achieved high yields for the assembled focal planes. Cold, dark characterization of the 280\,GHz focal planes was accomplished by taking IV curves at several bath temperatures ($T_{bath}$) from 461\,mK down to 330\,mK, the minimum temperature of the test dewar, with a gold-metalized silicon plate covering the feedhorns. Each curve provided a saturation power ($P$) for the detector at that temperature step. From there we obtained the critical temperature, conductivity of the bolometer legs at $T_c$ ($G_c$), and curvature ($\beta$) by fitting to the functional form 
\begin{equation}
P=\frac{G_0 T_0}{1+\beta} \left[ \left( \frac{T_c}{T_0}\right)^{1+\beta} - \left( \frac{T_{bath}}{T_0} \right)^{1+\beta} \right] \; ,
\end{equation}\label{eq:Pfit}
and extrapolating to 300\,mK, the nominal FPU operating temperature. The thermal conductance of the bolometer legs depends on the temperature like so
\begin{equation}
G(T) = G_0 \left( \frac{T}{T_0} \right)^{\beta} \; .
\end{equation}\label{eq:Gbeta}
Note that $T_0$ is the reference temperature where $G_0 \equiv G(T_0)$ is defined. Normal resistances ($R_n$) were obtained directly from the slope of the load curves at high bias. The resulting parameter estimates for all three 280\,GHz \spider focal planes are listed in Table~\ref{tab:params}. Figures~\ref{fig:phys_params}a and b show the saturation powers and critical temperatures across one 280\,GHz FPU. \\
\indent For all three arrays, the bolometer fabrication yield is $>$99\%. Additional loss of yield is either broken wiring or SQUID amplifier failures, the former being the dominant source. For example, the initial wire bonds could not withstand cold cycling due to differential thermal contraction between the detector array and the silicon breakout chips (Figure~\ref{fig:sau_skyin}). Optimizing bonding parameters appears to have largely solved the problem, but we suspect this remains the largest source of pixel loss.\\
\indent Based on load curve analysis at 330\,mK, the 280\,GHz FPUs have conservatively estimated end-to-end yields of 81\%, 88\%, and 95\% (Table~\ref{tab:params}), including wiring and readout defects. Subsequent fixes and warm continuity tests lead us to believe that these yields have improved further. 
\begin{table}[t!]
\begin{center}
\begin{tabular}{ c | c | c | c | c | c }
	 & \textbf{Target} & \textbf{Y3} & \textbf{Y4} & \textbf{Y5} & \textbf{Max. 1 $\sigma$}\\
	\hline\hline
	$P_{sat}$ & $<$3\,pW & 2.76\,pW & 2.79\,pW & 3.54\,pW & 6\%\\
	\hline
	$G_c$ & 30.7\,pW/K & 27.7\,pW/K & 27.0\,pW/K & 33.0\,pW/K & 6\%\\
	\hline
	$T_c$ & 420\,mK & 436\,mK & 445\,mK & 454\,mK & 2\%\\
	\hline
	$R_n$ & 11.5\,m$\Omega$ & 10.8\,m$\Omega$ & 10.7\,m$\Omega$ & 10.6\,m$\Omega$ & 4\%\\
	\hline
	$\beta$ & N/A & 1.88 & 1.96 & 2.03 & 6\%\\
	\hline\hline
	\textbf{Yield} &  & \textbf{81\%} & \textbf{88\%} & \textbf{95\%} & \\
\end{tabular}
\caption{\footnotesize End-to-end yield and parameter estimates at 300\,mK for the three 280\,GHz focal planes, dubbed Y3, Y4, and Y5 depending on which \textsc{Spider}-1 FPU their packaging came from. Parameters were measured at various bath temperatures down to 330\,mK, then extrapolated to the 300\,mK flight operating temperature (\S\ref{sec:characterization}). The median values are listed and the quoted spread at $1\sigma$ is the maximum for any of the FPUs.} \label{tab:params}
\end{center}
\vspace{-2.5em}
\end{table}
\vspace{-3mm}
\section{Conclusion and Status}\label{sec:conc}\vspace{-.5em}
The three 280\,GHz focal planes have been assembled and cryogenically screened. We find bolometer parameters $P_{sat}$, $T_c$, and $G_c$ that match our targets with high yield and uniformity ($<$6\% spread) across the wafer.
We are currently integrating the arrays into the telescopes and flight cryostat for further characterization. This will include polarimetric measurements, full-system optical efficiency and noise measurements, bandpass measurements via Fourier transform spectroscopy, and internal loading estimates.
The \textsc{Spider}-2 instrument is on track for a December 2018 flight from McMurdo Station in Antarctica. 

\vspace{-1mm}
\begin{acknowledgements} \textsc{Spider} is supported in the U.S. by the National Aeronautics and Space Administration under Grant No. NNX17AC55G and NNX12AE95G issued through the Science Mission Directorate and by the National Science Foundation through PLR-1043515. Additional support is provided by Department of Energy grant DE-SC007859. Corresponding author is supported by a National Science Foundation Graduate Research Fellowship.
Logistical support for the Antarctic deployment and operations was provided by the NSF through the U.S. Antarctic Program. Support in Canada is provided by the National Sciences and Engineering Council and the Canadian Space Agency. Support in Norway is provided by the Research Council of Norway.  Support in Sweden is provided by the Swedish Research Council through the Oskar Klein Centre (Contract No. 638-2013-8993).  We also wish to acknowledge the generous support of the David and Lucile Packard Foundation, which has been crucial to the success of the project. 
The collaboration is grateful to the British Antarctic Survey, particularly Sam Burrell, for invaluable assistance with data and payload recovery after the 2015 flight. 
\end{acknowledgements}

\pagebreak

\end{document}